\documentclass[preprint,showpacs,preprintnumbers,amsmath,amssymb,superscriptaddress]{revtex4}

\usepackage{graphicx}
\usepackage{dcolumn}
\usepackage{bm}
\usepackage{SIunits}
\usepackage{epstopdf}
\usepackage{array,color}
\usepackage{xspace}
\usepackage{CJK}

\begin{document}
\title{Behavior of vacuum and naked singularity under smooth gauge function in Lyra geometry}
\author{Haizhao Zhi}

\affiliation{Department of Physics, National University of Singapore, 117542, Singapore}

\date{\today}

\begin{abstract}
Lyra geometry is a conformal geometry originated from Weyl geometry.
In this article, we derive the exterior field equation under spherically symmetric gauge function $x^0(r)$ and metric in Lyra geometry. When we impose a specific form of the gauge function $x^0(r)$, the radial differential equation of the metric component $g_{00}$ will possess an irregular singular point(ISP) at $r=0$. 
Moreover, we apply the method of dominant balance and then get the asymptotic behavior of the new spacetime solution.
The significance of this work is that we could use a series of smooth gauge functions $x^0(r)$ to modulate the degree of divergence of the singularity at $r=0$ and the singularity will become a naked singularity under certain conditions. 
Furthermore, we investigate the physical meaning of this novel behavior of spacetime in Lyra geometry and find out that no spaceship with finite integrated acceleration could arrive at this singularity at $r=0$.
The physical meaning of gauge function and integrability is also discussed.

\end{abstract}


\email{h.zhi@u.nus.edu}
\vskip 300 pt

\maketitle

\section{Introduction}

Hermann Weyl\cite{weyl1918gravitation} created his version of generalized Riemannian geometry, Weyl geometry, to try to unify both gravitation and electromagnetism.   
The structure of Weyl geometry is insightful since it has a conformal structure which is an equivalent class of metric tensors $g$ under conformal transformation.
Nevertheless, the metric preserving character in Riemannian geometry is no longer established in Weyl geometry. (Unless the characteristic 1-form is exact, we could choose an effective metric to make the integrability being preserved within Weyl geometry\cite{Rosen1982weyl,Romero2012general,scholz2011weyl}.)
On the other hand, in 1951, Gerhard Lyra\cite{lyra1951modifikation} published 'Lyra geometry' as an alternative to Weyl geometry to solve the issue of non-metricity, in which he introduced another type of conformal transformation. 
In Lyra geometry, contracted curvature scalar shares similar expression as the curvature scalar calculated in Weyl geometry, which makes Lyra geometry another promising modified theory of gravity. (Note that $x^0$ is merely the symbol represents the gauge function following historical convention and has nothing to do with the time component of the spacetime coordinate.) 
Recently, many interesting theoretical topics in Lyra geometry have been widely discussed, like the singularity in a collapsing massive star\cite{ziaie2014trapped}, cosmology models in Lyra geometry\cite{shchigolev2016cosmology,shchigolev2016inhomogeneous,shchigolev2013on,singh1997new,singh2017could,reddy2017bianchi}, gauge field as the source of gravity\cite{shchigolev2017exact,pradhan2009inhomogeneous,ali2014invariant,zhi2014new}, so on and so forth.
In this article, after we choose a series of smooth and spherically symmetric gauge functions, singularities emerge at $r=0$. 
Moreover, the degree of divergence of these singularities could be modulated if we choose different smooth gauge functions $x^0$. 
Under specific mathematical conditions, the singularities at $r=0$ are no longer hiding behind an event horizon and becoming naked singularities. 
Historically, with some specific conditions, naked singularity could be formulated in Einstein's general relativity. 
For instance, M. Choptuik et al.\cite{choptuik1996critical} have found that the existence of solution of naked singularity is on a 'verge point' relative to black hole in the phase diagram through numerical investigation.
In addition, recently, T. Crisford and J. E. Santos\cite{crisford2017violating} propose a model in which naked singularity could be found embedded in a saddle-shaped geometry in four dimensional anti-de Sitter space. 
Although, in their discovery, if another type of force presents in that universe which affects particles more strongly than gravity, the naked singularity will be cloaked by an event horizon.
The idea that naked singularities are forbidden from forming in nature originates from Roger Penrose\cite{hawking1970singularities,penrose2002golden,wald1999gravitational}, which is called the cosmic censorship conjecture. 
Furthermore, we could raise another question that could cosmic censorship conjecture be observationally tested? 
K.S. Virbhadra and G.F.R. Ellis\cite{virbhadra2002gravitational,virbhadra2000schwarzschild} have shown that black holes and naked singularities can be observationally differentiated through their gravitational lensing characteristics. 
Gravitational lensing is an important astrophysical tool for observationally testing the cosmic censorship conjecture. 
For example, a black hole and a naked singularity of the same ADM mass with same symmetry, acting as gravitational lenses, will render different number and different orientations of images of the same light source.
The main purpose of this paper is to expand the solution set of spacetime metrics in Lyra geometry and explain the physical implication of it. 
What is more, with the naked singularity being presented within Lyra geometry, we could prove that a spaceship will never reach infinitely small radial coordinate $r$ under this spacetime solution with finite integrated acceleration\cite{chakrabarti1983timelike}.
This article is a trial of a rigorous discussion of using possible applications of the analytic approximation methods in solving the spherical metric model in Lyra geometry.

\section{The key concepts in Lyra geometry}

Weyl geometry has a conformal structure, which is physically intuitive and similar to $U(1)$ gauge symmetry. Instead of scaling the metric in Weyl geometry, G. Lyra promoted the concept of scaling to the vector basis on manifold.
The basis in the tangent space $T_m(M)$ at point $m$ on Lyra manifold $M$ is defined as $\tilde{e}_{\mu}(m)= \{ \frac{1}{x^0(x^{\mu})} \frac{\partial}{\partial x^{\mu}} \}$ and the basis in the cotangent space $[T_m(M)]^{*}$ is defined as $\tilde{e}^{\mu}(m)={x^0(x^{\mu})}\mathrm{d}x^{\mu}$, with $\mu = 0, 1, 2, 3$. 
Here ${x^0(x^{\mu})}$ is a nonzero and smooth gauge function defined in Lyra geometry. 
Moreover, a reference system in Lyra geometry can be written as: ($x^0|x^{\mu}$), which is a combination of a coordinate system $\{x^{\mu} \}$ and a gauge function ${x^0(x^{\mu})}$.

The basis in the tangent space $\tilde{e}_{\mu}$ under a transformation of the reference system can be expressed as $\tilde{e}_{\mu{'}}=\lambda^{-1} A^{{\mu}}_{\mu{'}} \tilde{e}_{\mu}$.
In addition, the vector components under a transformation of the reference system can be written as $\phi^{{\mu}{'}}=\lambda A^{{\mu}{'}}_{\mu}\phi^{\mu}$, where $A^{{\mu}{'}}_{\mu}=\frac{\partial x^\mu{'}}{\partial x^{\mu}}$ is the Jacobian in Riemannian fashion, with det$A^{{\mu}{'}}_{\mu} \neq 0$ and $\lambda={x^{0'}}/x^0$. 
Here we could observe that $\phi^{\mu}\tilde{e}_{\mu}$ has the same 'physical' meaning as a vector $\phi$(real vector, not vector components) in Riemannian geometry. 
A vector(or any tensorial quantities) in Lyra geometry is invariant under any transformation of the reference system, same as Riemannian geometry. Now we could dig this concept one step further: in Riemannian geometry, metric tensor: $g_{\mu\nu}\mathrm{d}x^{\mu}\mathrm{d}x^{\nu}$ and 'line element': $\mathrm{d}s^2$ are interchangeable physical quantities.
Here, in Lyra geometry, the metric tensor(line element) could be expressed as $\mathrm{d}s^2=(x^0)^2g_{\alpha\beta}\mathrm{d}x^{\alpha}\mathrm{d}x^{\beta}$, where $g_{\alpha\beta}=g(\tilde{e}_\alpha,\tilde{e}_\beta)=\mathrm{d}s^2(\tilde{e}_\alpha,\tilde{e}_\beta)$; the expression of the line element $\mathrm{d}s^2$ is invariant under the transformation of the reference system.

\subsection{Connection in Lyra geometry}

D. K. Sen et al. has proved that\cite{sen1972weyl} a connection is uniquely defined on a manifold $M$ if we have: 
\begin{equation}\label{C.C.1}
    (\nabla_Zg)(X,Y) = A(Z,X,Y) 
\end{equation}
\begin{equation}\label{C.C.2}
    Tor_{\nabla}(X,Y) = B(X,Y)
\end{equation}
where $A(Z,X,Y)$ is symmetric with respect to the two smooth vector fields $X$ and $Y$ and $B(X,Y)$ is antisymmetric with respect to $X$ and $Y$($Z$ is also a smooth vector field.).
Specifically, when $A(Z,X,Y)=-\phi(Z) g(X,Y)$ and $ Tor_{\nabla}(X,Y)=0$, the geometry is Weyl geometry, where $\phi \in \Lambda^1(M)$ is a smooth global one-form field on $M$.

D. K. Sen et al. also proved that\cite{sen1972weyl} the connection on the manifold $M$ is uniquely defined following the expression below: 
\begin{small}
\begin{equation}\label{C.C.3}
\begin{split}
2g(\nabla_Z,Y)=X(g(Y,Z))+Y(g(X,Z))-Z(g(X,Y))-A(X,Y,Z)-A(Y,X,Z)+A(Z,X,Y)-\\
g(B(X,Z),Y)-g(B(Y,X),Z)+g(B(Z,Y),X)-g([X,Z],Y)-g([Y,X],Z)+g([Z,Y],X)
\end{split}
\end{equation}
\end{small}

Here, Lyra geometry is defined using a smooth 1-form field $\phi$ with $A(X,Y,Z) = 0$ and $ B(X,Y) =\frac{1}{2}[\phi(Y)X - \phi(X)Y]$. Notice that the Lie brackets are no longer zero since the basis vector now coupled with the gauge function $x^0(x^{\mu})$: $[\tilde{e}_\alpha,\tilde{e}_\beta]=\frac{1}{2}(\delta^{\mu}_{\alpha}\mathring{\phi}_{\beta}-\delta^{\mu}_{\beta}\mathring{\phi}_{\alpha})\tilde{e}_\mu$, where $\mathring{\phi}_{\alpha}$ is not vector components.
$\mathring{\phi}_{\alpha}$ is merely a symbol of four fold functions: $\mathring{\phi}_{\alpha}=-2\partial_{\alpha}(1/x^0)$.

From now on, we will call $g_{\alpha\beta}$ as 'metric' and $(x^0)^2g_{\alpha\beta}$ as 'usual metric' following Sen's convention in (2.22) in \cite{sen1971scalar}. The component of the connection of Lyra geometry is: $\nabla_{\tilde{e}_\nu}{\tilde{e}_\mu}=\tilde{\Gamma}^{\gamma}_{\mu\nu}\tilde{e}_{\gamma}$. Plug $X=\tilde{e}_{\mu}$, $Y=\tilde{e}_{\nu}$, $Z=\tilde{e}_{\gamma}$ into eq.(\ref{C.C.3}), we could derive the Lyra connection $\tilde{\Gamma}^{\alpha}_{\mu\nu}$ in the component form:

\begin{equation}\label{C.L.6}
\tilde{\Gamma}^{\alpha}_{\mu\nu}=\frac{1}{x^0}\Gamma^{\alpha}_{\mu\nu}+S^{\alpha}_{\mu\nu}
\end{equation}

\begin{equation}\label{C.L.7}
S^{\alpha}_{\mu\nu}=\frac{1}{2}(\delta^{\alpha}_{\nu}\phi_{\mu}-g_{\nu\mu}\phi^{\alpha}),
\end{equation}
where $\Gamma^{\alpha}_{\mu\nu}$ is the Christoffel symbol in Riemannian fashion. $\phi_\alpha$ here is defined as: $\phi_\alpha=\phi(\tilde{e}_\alpha)+\mathring{\phi}_\alpha$ and $\phi^{\alpha}=g^{\alpha\mu}\phi_{\alpha}$.
We need to note that $\phi_\alpha$ is not components of a dual vector field $\phi$. 
It is a 'mixture' of vector components and four fold functions. 
When we perform coordinate transformation, $\phi_{\alpha}$ defined in eq.(\ref{C.L.7}) is not going to transform as components of dual vector field as: $\phi_{\mu{'}}=\lambda^{-1} A^{{\mu}}_{\mu{'}} \phi_{\mu}$ above\footnote{The formula of $\phi_{\alpha}$ under coordinate transformation is given in reference\cite{sen1971scalar,sen1972weyl}:  $\phi_{\alpha} \to \phi_{\alpha^{'}}= \lambda^{-1}A^{\alpha}_{\alpha^{'}}[\phi_{\alpha}+ (x^0)^{-1}\partial_{\alpha}(ln\lambda^2)]$, following the transformation law of $\tilde{\Gamma}^{\mu}_{\nu\lambda}$.}.
This renders the physics of gravitational theory in Lyra geometry is dependent on $x^0$ of the reference system.
Moreover, we could observe that $\tilde{\Gamma}^{\alpha}_{\mu\nu}$ is no longer symmetric with respect to ${\mu}$ and ${\nu}$ since it is not torsion free\footnote{There are two minor typos in the reference\cite{sen1971scalar}, one is below eq.(1.12): $ \dots {\phi}_{\lambda{'}}={\frac{1}{x^{0'}}}{\frac{\partial ln{{\lambda}^2}}{\partial x^{\lambda{'}}}}$, the author missed one prime on $x^0$ in the denominator. The other one is in eq.(3.19): $'+'$ is actually $'-'$.}.

\subsection{Physical discussion of gauge function and integrability}


In Weyl geometry, $\nabla_{\alpha}g_{\mu\nu}=\phi_{\alpha}g_{\mu\nu}$, component form of $(\nabla_Zg)(X,Y)=-\phi(Z)g(X,Y)$, is called the Weyl condition of compatibility or the non-metricity condition, since the metric is not preserved under parallel shifting.
Einstein's critique about Weyl geometry is that if $\phi$ is a pure geometric object, then the existence of sharp spectral lines in atomic physics would be no longer possible since the physics would depend on their past history\cite{pauli1981theory}.
This issue could be resolved that, if we impose $\phi = \mathrm{d}f$, where $f$ is a smooth scalar function defined on $M$, and redefine the metric as $\hat{g}=e^{-f}g$, we will have: $\nabla_{\alpha}\hat{g}_{\mu\nu}=0$. 
The integration of the metric tensor $\hat{g}$ will no longer be path dependent \cite{Romero2012general}. 
We could replace the metric $g_{\mu\nu}$ by $\hat{g}_{\mu\nu}$ and the invariance of the gauge transformation will be preserved, then the line element in integrable Weyl geometry is $\mathrm{d}s^2=e^{-f}g_{\mu\nu}\mathrm{d}x^{\mu}\mathrm{d}x^{\nu}$.
On the other hand, Lyra geometry is defined by $(\nabla_Zg)(X,Y)=0$ and $ Tor_{\nabla}(X,Y)=\frac{1}{2}[\phi(Y)X - \phi(X)Y]$. 
The metric in Lyra geometry is preserved under parallel shifting because of $(\nabla_Zg)(X,Y)=0$, then Lyra geometry is a naturally integrable geometry comparing to Weyl geometry. 
One thing need to be discussed here is that the basis vector defined in the cotangent space  $[T_m(M)]^{*}$ is '$x^0 \mathrm{d}x^{\mu}$' instead of '$\mathrm{d}x^{\mu}$' so that the line element in Lyra geometry is $\mathrm{d}s^2=(x^0)^2g_{\mu\nu}\mathrm{d}x^{\mu}\mathrm{d}x^{\nu}$.
For instance, we have a rod like object with $1cm$ length measured at the spacelike hypersurface at the starting point $a$ and then parallel shift the object to point $b$. We assume that the spacetime manifold is flat and static and the gauge function is smoothly changing from $x^0=1$ at point $a$ to $x^0=2$ at point $b$. 
Here, at point $b$, the length of the rod like object is still $1cm$ if you use the ruler at point $b$, but the length of the ruler itself is twice prolonged.
This is the physical meaning of integrability in the context of Lyra geometry.
We could observe that, although the mathematical starting point is different, if we assume that $x^0= e^{-\frac{f}{2}}$, Lyra geometry can be tuned to be equivalent to an integrable Weyl geometry locally so that Lyra geometry can also be treated as a scalar-tensor theory\cite{sen1971scalar,Romero2012general}.
As a result, Lyra geometry has both the conformal structure as Weyl geometry and integrability; nevertheless, the tradeoff is that there is no gauge transformation symmetry in Lyra geometry.

\subsection{Curvature and field equation in Lyra geometry}

Curvature tensor in Lyra geometry is defined in a same fashion as Riemannian geometry, which is a map $K$: $V(M) \otimes V(M) \otimes V(M) \rightarrow V(M)$. The component of the curvature tensor could be given as\cite{sen1971scalar}:
\begin{equation}\label{CU.1}
    K^{\mu}_{\lambda\alpha\beta}=(x^0)^{-2}[\frac{\partial(x^0\tilde{\Gamma}^{\mu}_{\lambda\beta})}{\partial x^{\alpha}}-\frac{\partial(x^0\tilde{\Gamma}^{\mu}_{\lambda\alpha})}{\partial x^{\beta}}]+\tilde{\Gamma}^{\mu}_{\rho\alpha}\tilde{\Gamma}^{\rho}_{\lambda\beta}-\tilde{\Gamma}^{\mu}_{\rho\beta}\tilde{\Gamma}^{\rho}_{\lambda\alpha}
\end{equation}

Moreover, we could do the contraction to get the Ricci tensor and curvature scalar in Lyra's fashion: $K_{\alpha\beta}=K^{\mu}_{\alpha\beta\mu}$ and $K=g^{\alpha\beta}K_{\alpha\beta}$, where 
\begin{equation}\label{CU.1.5}
K=(x^0)^{-2}R+3(x^0)^{-1}\nabla_{\alpha}\phi^{\alpha}+\frac{3}{2}\phi_{\alpha}\phi^{\alpha}
\end{equation}
To get the gravitational field equation in Lyra geometry, we need to apply the variational principle on the curvature scalar as : $\delta\int K(-g)^{\frac{1}{2}} x^0 \mathrm{d}x^1 \cdots x^0 \mathrm{d}x^4=0$, where $g$ is the determinant of the metric $g_{\alpha\beta}$.
The variational operator $\delta$ is commutable with the gauge function $x^0$. 
We will have two equations each corresponds to $g_{\alpha\beta}$ and $\phi^{\alpha}$ after  applying the variational principle on the curvature scalar.
The final expression is\cite{sen1971scalar}: 
\begin{small}
\begin{equation}\label{CU.2}
\begin{split}
\int \{ -[R^{\alpha\beta}-\frac{1}{2}g^{\alpha\beta}R+\frac{3}{2}(x^0)^2\phi^{\alpha}\phi^{\beta}-\frac{3}{4}(x^0)^2g^{\alpha\beta}\phi_{\nu}\phi^{\nu}
-\frac{3}{4}(x^0)^2g^{\alpha\beta}\mathring{\phi}_{\nu}\phi^{\nu}\\
+\frac{3}{2}(x^0)^2\mathring{\phi}^{\alpha}\phi^{\beta}](x^0)^2(-g)^{\frac{1}{2}}\delta g_{\alpha\beta} \}\mathrm{d}x^1 \cdots \mathrm{d}x^4 =0
\end{split}
\end{equation}
\end{small}
with respect to $g_{\alpha\beta}$ and

\begin{equation}\label{CU.3}
    \int \{ [3\phi^{\alpha}+\frac{3}{2}\mathring{\phi}^{\alpha}](x^0)^4(-g)^{\frac{1}{2}}\delta\phi_\alpha \}\mathrm{d}x^1 \!\cdots \!\mathrm{d}x^4 =0
\end{equation}
with respect to $\phi^{\alpha}$.

Furthermore, we will reach at two exterior field equations:
\begin{equation}\label{CU.4}
    R^{\alpha\beta}-\frac{1}{2}g^{\alpha\beta}R+\frac{3}{2}(x^0)^2\phi^{\alpha}\phi^{\beta}-\frac{3}{4}(x^0)^2g^{\alpha\beta}\phi_{\nu}\phi^{\nu}-\frac{3}{4}(x^0)^2g^{\alpha\beta}\mathring{\phi}_{\nu}\phi^{\nu}+\frac{3}{2}(x^0)^2\mathring{\phi}^{\alpha}\phi^{\beta}=0
\end{equation}
\begin{equation}\label{CU.5}
    3\phi^{\alpha}+\frac{3}{2}\mathring{\phi}^{\alpha}=0
\end{equation}
Here, from eq.(\ref{CU.5}) and eq.(\ref{CU.4}), we could observe that our theory is a special case of the interior field equations of Brans and Dicke theory\cite{brans1961mach}, where the Brans-Dicke constant is $\omega=\frac{3}{2}$.
We could plug eq.(\ref{CU.5}) into eq.(\ref{CU.4}) to get the final exterior field equation. 
\begin{equation}\label{CU.6}
    R_{\alpha\beta}-\frac{1}{2}g_{\alpha\beta}R-\frac{3}{2}(x^{0})^2 x^0,_{\alpha}x^0,_{\beta}+\frac{3}{4}(x^0)^{-2}g_{\alpha\beta}x^0,_{\nu}x^{0{,\nu}} = 0
\end{equation}
We could see that, although $\phi_{\alpha}$ is defined as a 'mixture' of vector components and gauge function in eq.(\ref{C.L.7}), the variational principle informs us that gauge function $x^0$ is the only quantity we need to define a Lyra geometry.
On the other hand, like in Riemannian general relativity, the interior field equation could be formulated as\cite{sen1971scalar}:
\begin{equation}\label{CU.7}
    R_{\alpha\beta}-\frac{1}{2}g_{\alpha\beta}R-\frac{3}{2}(x^0)^2x^0,_{\alpha}x^0,_{\beta}+\frac{3}{4}(x^0)^{-2}g_{\alpha\beta}x^0,_{\nu}x^{0{,\nu}} = -[{8 \pi G}/(x^0)^2]T_{\alpha\beta}
\end{equation}

There is one conundrum about the normal gauge($x^0 \equiv 1$ globally on the manifold) in Lyra geometry worth discussing here. In many of the literature before, authors who are using the concept of normal gauge tend to plug $x^0 =1 $ into eq.(\ref{CU.1.5}) and get $K=R+3\nabla_{\alpha}\phi^{\alpha}+\frac{3}{2}\phi_{\alpha}\phi^{\alpha}$ and then do the variation on $K$. In this scenario, the smooth 1-form field $\phi$ remains in the gravitational field equation. The possible drawback of this operation is that the meaning of the 1-form field $\phi$ is unclear. On the other hand, if we directly do the variational operation on $K$ in eq.(\ref{CU.1.5}), we could observe that the components of the smooth 1-form field $\phi$ will be fully determined by $x^0$ in eq.(\ref{CU.5}) so that physics is determined only by the gauge function. If we choose the normal gauge condition $x^0 \equiv 1$ here, the theory will be the same as general relativity. We are going to use the latter scenario here.

In next section, we will explore more about the spherical solution of the exterior field equation (\ref{CU.6}) and the behavior near the origin point with $r \to 0^{+}$ in our spacetime coordinate system.


\section{The Exterior Field Equation}

\subsection{The exterior field equation in Lyra geometry}

The vacuum field equation now is written as\cite{sen1971scalar,halford1972scalar}:
\begin{equation}\label{E.1}
    R_{\alpha\beta}-\frac{1}{2}g_{\alpha\beta}R-\frac{3}{2}(x^0)^2x^0,_{\alpha}x^0,_{\beta}+\frac{3}{4}(x^0)^{-2}g_{\alpha\beta}x^0,_{\nu}x^{0{,\nu}} = 0
\end{equation}
Mathematically, the $R_{\alpha\beta}$ and $R$ in eq.(\ref{E.1}) have the same mathematical expression as Ricci Tensor and Ricci Scalar in Riemannian geometry.
The $x^0$ here is the non-zero gauge function defined on a specific patch of the atlas defined on the manifold($x^0,_{\nu}$ means $\partial x^0/ \partial x^{\nu}$.).

\subsection{Static and spherically symmetric solution}

Here, we impose the spherical and static gauge function: $x^0=x^0(r)$ on the spacetime in Lyra geometry.
The spherical metric can be expressed as:
    \begin{equation}\label{E.2}
        g_{\alpha\beta}=\begin{bmatrix}
        -e^{\nu}&0&0&0\\
0&e^{\lambda}&0&0\\
0&0&r^2&0\\
0&0&0&r^2sin^2\theta
\end{bmatrix},
    \end{equation}
where $\lambda=\lambda(r)$ and $\nu=\nu(r)$, both are spherical and time invariant functions.
Plug this metric back into the field equation (\ref{E.1}), we will have a set of differential equations of the components of this spherical metric. 
After lengthy calculation, the set of the differential equations of the metric components is:

\begin{equation}\label{component1}
{\nu}^{'}/r+(1-e^{\lambda})/r^2+\frac{3}{4}f(r)=0     
\end{equation}

\begin{equation}\label{component2}
-r^2 e^{-\lambda} [\nu^{{'}{'}}/2-\lambda^{'}\nu{'}/4+\nu^{'2}/4+(\nu^{'}-\lambda^{'})/(2r)]+\frac{3}{4}r^2e^{\nu-\lambda}f(r)=0
\end{equation}

\begin{equation}\label{component3}
e^{\nu-\lambda}[-\lambda^{'}/r+(1-e^{\lambda})/r^2]-\frac{3}{4}e^{\nu-\lambda}f(r)=0
\end{equation}
The definition of $f(r)$ within these equations is : $f(r)=[x^{0'}(r)/x^0(r)]^2$, in which the prime means taking derivative with respect to the radial coordinate $r$.

From the results above, we could see, if we cancel $e^{\nu-\lambda}$ factor in equation (\ref{component3}) and then directly use equation (\ref{component1}) minus equation (\ref{component3}), we will have a relationship between $\nu$, $\lambda$ and $f(r)$ as:
      \begin{equation}\label{component4}
      \nu^{'}+\lambda^{'}+\frac{3}{2}rf(r)=0
      \end{equation}
This process is similar to the calculation of the Schwarzschild solution in general relativity. 
We could observe that if we know the solution of $\nu(r)$, we definitely will be able to solve $\lambda(r)$, since equation (\ref{component4}) will become a first order linear ordinary differential equation(ODE).

Plug (\ref{component4}) into (\ref{component2}), we will get:
      \begin{equation}\label{E.3}
{\nu}^{{'}{'}}+\frac{2}{r}\nu^{'}+\nu^{'2}+\frac{3}{4}\nu^{'}rf(r)=0
      \end{equation}
We could observe that this is a second order nonlinear differential equation.
To make the differential equation a linear ODE, we could see that ${\nu}^{''}+\nu^{'2}$ has similar form as the second derivative of $e^{\nu}$ divide itself, which is: $(e^{{\nu}}){{'}{'}}=\nu^{''}e^{\nu}+(\nu^{'})^2e^{\nu}$. 
We then assume that $y=e^{\nu}$, plug it back to eq.(\ref{E.3}), we now have a linear ODE:
\begin{equation}\label{E.4}
y^{{'}{'}}+y^{'}(\frac{2}{r}+\frac{3}{4}rf(r))=0
\end{equation}
where $y$ is $e^{\nu(r)}$. Here $-e^{\nu(r)}$ is the $g_{00}$ component of the metric in eq.(\ref{E.2}) and $\nu(r)$ is the radial function appeared in eq.(\ref{component1}, \ref{component2}, \ref{component3}).

For instance, if we choose the normal gauge, which is $x^0=1$ on one specific patch on the manifold. We will have $f(r)=[x^{0'}(r)/x^0(r)]^2=0$, which renders the eq.(\ref{E.4}) as:
      \begin{equation}\label{E.5}
          y^{{'}{'}}+y^{'}\frac{2}{r}=0
      \end{equation}
Since this ODE is directly integrable, we could calculate the solution of it: $y=-(C_2+C_1 \cdot r)/r$, where $C_1$ and $C_2$ are the constants of integration.

Recall that $y=e^{\nu}=-g_{00}$, from the Newtonian approximation, we have $g_{00}=-(1-\frac{2M}{r})$ as our first condition. In addition, when the radial coordinate $r$ is approaching to $\infty$, the time component of our metric $g_{00}$ will be approaching $-1$, since the spacetime will be gradually flattened. This is the boundary condition of the metric.
In the end, we could have two restrictions on the constants of integration: $C_1=-1, C_2=2M$, which means that the choice of normal gauge $x^0=1$ has completely restored the Schwarzschild solution in general relativity.


\section{Tuning The Singularity At $r=0$}

Now we would like to apply a non-trivial gauge function $x^0(r)$.
Here we could raise one question: how different levels of divergence in $f(r)$ at $r=0$ would influence the behavior of the metric $g_{\mu\nu}$ near $r=0$? (We need to notice that $f(r)$ is divergent as the radial coordinate $r$ approaches to $0$ does NOT mean the gauge function $x^0(r)$ is also divergent as $r \to 0$.)

\subsection{Choosing a specific $x^0(r)$}

From now on, we will be focusing on a specifically interesting situation which is $f(r)=\frac{A}{r^{2+\alpha}}$, where $\alpha \in \mathbb{R}^{+}$ and $A \in \mathbb{R}^{+}$\footnote{If $A \leq 0$, it can be shown that the gauge function $x^0$ will be imaginary. Since $x^0$ is the scaling factor of the basis vector in the cotangent space, $x^0$ must be positive.}. The reason that we choose this expression for $f(r)$ is that the radial differential equation (\ref{E.4}) would have an irregular singular point(ISP) at $r=0$ and the spherically symmetric gauge function $x^0(r)$ will remain smooth in $r \in [0,+\infty)$ at the same time. Surprisingly, we could further see that, not only the singularity will be induced at $r=0$ in the usual metric we solved, the event horizon which is deemed to emerge and protect the singularity is actually disappeared when $\alpha \in (\sqrt{2}, \infty)$. Theoretically, the existence of naked singularity plays a key role in the cosmic censorship conjecture. This paper has mathematically proved that naked singularity can be introduced in Lyra geometry with smooth gauge function being applied.

Since $f(r)=\frac{A}{r^{2+\alpha}}$, from the definition of $f(r)$ which is $f(r)=[x^{0'}(r)/x^0(r)]^2$, we can derive that:
\begin{equation}\label{T.1}
\frac{x^{0'}(r)}{x^0(r)}=\frac{\sqrt{A}}{r^{1+\alpha/2}}
\end{equation}
by taking square root of $f(r)$. Because eq.(\ref{T.1}) is a first order ODE of the gauge function, we could explicitly solve this spherically symmetric gauge function $x^0(r)$ as:
\begin{equation}\label{T.2}
x^0(r)=C' \cdot exp(\int \frac{\sqrt{A}}{r^{1+\alpha/2}} \mathrm{d}r) = C \cdot exp(-\frac{2\sqrt{A}}{\alpha r^{\alpha/2}})
\end{equation}
Here $C'$ and $C$ are two nonzero constants of integration. No matter how we choose $C$, bigger or smaller, the final expression of the ODE (\ref{E.4}) and the metric in eq.(\ref{E.2}) are not changed.

Moreover, we could observe that as $r \to \infty$, $x^0(r)$ will be approaching to $C$ and as $r \to 0^{+}$,  $x^0(r)$ is approaching to $0$. Furthermore, we could show that this function is a smooth function in $r \in (0,\infty)$ and the right derivative taken at $r=0^+$ approaches to zero. In the meantime, the linear ODE of eq.(\ref{E.4}) has an ISP at $r=0$. We will use the dominant balance method\cite{bender2013advanced} to determine the asymptotic behavior of eq.(\ref{E.4}) and calculate the asymptotic solution of the usual metric of this model when the radial coordinate $r$ is approaching to $0^{+}$.

\subsection{Dominant balance at the irregular singular point (ISP) $r=0$}

Now we consider eq.(\ref{E.4}) in the last section, within this ODE, $y=e^{\nu(r)}=-g_{00}$. Here, we plug $f(r)=\frac{A}{r^{2+\alpha}}$ into eq.(\ref{E.4}), then this is corresponding to the situation where the gauge function is $x^0(r) = C \cdot exp(-\frac{2\sqrt{A}}{\alpha r^{\alpha/2}})$ as we derived in eq.(\ref{T.2}). Explicitly, we will have:
\begin{equation}\label{I.1}
y^{{'}{'}}+\frac{2}{r}y^{'}+\frac{\frac{3}{4} A}{r^{1+\alpha}}y^{'}=0
\end{equation}
We are going to apply dominant balance method onto eq.(\ref{I.1}). Moreover, the method of dominant balance will be briefly introduced here:

For a regular ODE, like:
\begin{equation}\label{I.3}
\Psi^{''}+\frac{p(x)}{x-x_0}\Psi^{'}+\frac{q(x)}{(x-x_0)^2}\Psi=0,
\end{equation}
where $\Psi(x)$, $p(x)$ and $q(x)$ are real-valued functions defined in $\mathbb{R}$. If this ODE has a regular singular point at $x=x_0$, which means $p(x)$ and $q(x)$ are both analytic at $x=x_0$, we could explicitly solve this ODE using standard Frobenius method. On the other hand, what if the functional coefficients in eq.(\ref{I.3}), $\frac{p(x)}{x-x_0}$ and $\frac{q(x)}{(x-x_0)^2}$, are divergent at $x=x_0$? In this scenario, we often could NOT get the explicit analytic solution of this ODE. Nevertheless, we could solve the asymptotic behavior near $x=x_0$ using the dominant balance method. The procedure of dominant balance method is like this:
\begin{itemize}
\item $\mathbf{1}$. Replace $\Psi(x)$ by $\Psi(x)=e^S(x)$, where $S(x)$ is also a real-valued function defined in $\mathbb{R}$. There will be several terms involving $S$, $S^{'}$ and $S^{{'}{'}}$.
\item $\mathbf{2}$. We assume any two of the terms to be the 'dominant' ones, when $x$ is approaching to $x_0$. Omit the other terms and solve the asymptotic behavior of the remaining ODE of $S(x)$ with the two 'dominant' terms.
\item $\mathbf{3}$. If these two terms are really 'dominant', which means the ratio of the two chosen terms to any other terms we just omitted is approaching to infinite as $x$ is approaching to $x_0$, we will have the real asymptotic behavior of $S(x)$. Otherwise, we choose two other terms and redo the step 2.
\end{itemize}
To get the asymptotic behavior near the ISP at $r=0$, we plug $y=e^S$ into equation (\ref{I.1}), then equation (\ref{I.1}) becomes:
\begin{equation}\label{I.4}
    (S^{'})^2+S^{{'}{'}}+\frac{2}{r}S^{'}+\frac{\frac{3}{4} A}{r^{1+\alpha}}S^{'}=0
\end{equation}
After different trials, we assume the first and the fourth terms are the 'dominant' ones, then we have:
\begin{equation}\label{I.5}
    (S^{'})^2 \sim -S^{'}\frac{\frac{3}{4}A}{r^{1+\alpha}}
\end{equation}

Following the method of dominant balance, the ratio of these two chosen terms to the other terms needs to approach to infinity. First, we could solve $S^{'}$ in eq.(\ref{I.5}), the result is:
\begin{equation}\label{I.6}
    S^{'} \sim -\frac{\frac{3}{4}A}{r^{1+\alpha}}
\end{equation}
Taking derivative with respect to $r$ on both sides in eq.(\ref{I.6}), we could conclude that:
\begin{equation}\label{I.7}
    S^{{'}{'}} \sim \frac{\frac{3}{4}A(1+\alpha)}{r^{2+\alpha}} \ll (S^{'})^2
\end{equation}
So, from eq.(\ref{I.7}), the second term in eq.(\ref{I.4}) is negligible compare to the first and the fourth terms, when $r \to 0$. As to the third term in eq.(\ref{I.4}), the relationship between it and the two 'dominant' ones is:
\begin{equation}\label{I.8}
\frac{2}{r}S^{'} \ll (S^{'})^2 \sim -S^{'}\frac{\frac{3}{4}A}{r^{1+\alpha}}
\end{equation}
We could conclude that, as $\alpha > 0$ and $r \to 0$, the asymptotic behavior of $S$ is the solution of eq.(\ref{I.6}) after we replace the '$\sim$' to '$=$'. The solution of $S$ is: 
\begin{equation}\label{I.9}
S \sim \frac{\frac{3}{4}A{\alpha}}{r^{\alpha}},
\end{equation}
where we omit the constant of integration since $S$ is divergent when $r \to 0$. Next, we could plug eq.(\ref{I.9}) into eq.(\ref{I.1}) to solve the asymptotic behavior of $y$ in the initial radial ODE. The asymptotic behavior of the eq.(\ref{I.1}) is:
\begin{equation}\label{I.10}
y = - g_{00} \sim e^S = exp(\frac{\frac{3}{4}A{\alpha}}{r^{\alpha}})
\end{equation}
under the condition that $\alpha > 0$ and $r \to 0$.

In conclusion, since $-g_{00}=e^{\nu(r)}$ $\sim$ $exp(\frac{\frac{3}{4}A{\alpha}}{r^{\alpha}})$, we know that the asymptotic behavior of the spherical function $\nu(r)$ is $\nu(r)$ $\sim$ ${\frac{3}{4}A{\alpha}}/{r^{\alpha}}$.
Moreover, we could solve the asymptotic behavior of the spherical function $\lambda(r)$ 
using eq.(\ref{component4}) by pluging $\nu(r)$ $\sim$ ${\frac{3}{4}A{\alpha}}/{r^{\alpha}}$ and $f(r)=\frac{A}{r^{2+\alpha}}$ into it. We will have:
\begin{equation}\label{Lambda.1}
    \lambda^{'}=-\frac{3}{2}r\frac{A}{r^{2+\alpha}}+\frac{3}{4}A{\alpha}^2\frac{1}{r^{1+\alpha}}=\frac{1}{r^{1+\alpha}}(\frac{3}{4}A\alpha^2-\frac{3}{2}A)
\end{equation}
Furthermore, we perform integration on eq.(\ref{Lambda.1}) to derive the explicit asymptotic form of $\lambda(r)$:
\begin{equation}\label{Lambda.2}
    \int \lambda^{'}\mathrm{d}r=\frac{3}{4}A(\alpha^2-2)(\frac{-\alpha}{r^{\alpha}}+C^{''}),
\end{equation}
where $C^{''}$ is the constant of integration.
So the $g_{11}$ component in eq.(\ref{E.2}) is: $g_{11}=exp({\frac{3}{4}A(\alpha^2-2)(\frac{-\alpha}{r^{\alpha}}+C^{''})})$. For spherical metric, when $r \to \infty$, $g_{11}$ needs to be approaching to 1. Since $\alpha > 0$, if $\alpha^2 \neq 2$, $C^{''}$ must be chosen as 0. Otherwise, when $\alpha^2=2$, $g_{11}$ will be equal to 1 no matter how we choose the integration constant.
Here, $g_{11}$ could be written as $g_{11}=exp({\frac{3}{4}A(\alpha^2-2)\frac{-\alpha}{r^{\alpha}})}$. We will discuss the physical meaning of $g_{11}$ later in this section.

In the next step, we plug both eq.(\ref{I.10}) and eq.(\ref{Lambda.2}) into eq.(\ref{E.2}), then we will have the explicit asymptotic expression for the spacetime metric:
   \begin{equation}\label{Metric.1}
        g_{\alpha\beta}=\begin{bmatrix}
        -exp(\frac{\frac{3}{4}A{\alpha}}{r^{\alpha}})&0&0&0\\
0&exp(-{\frac{3}{4}A(\alpha^2-2)\frac{\alpha}{r^{\alpha}}})&0&0\\
0&0&r^2&0\\
0&0&0&r^2sin^2\theta
\end{bmatrix}
    \end{equation}
In Lyra's geometry, the line element is given as\cite{sen1971scalar}:
\begin{equation}\label{Metric.2}
\mathrm{d}s^2=(x^0)^2g_{\alpha\beta}\mathrm{d}x^{\alpha}\mathrm{d}x^{\beta},
\end{equation}
since the basis in $[T_m(M)]^{*}$ is defined as $\tilde{e}^{\mu}(m)={x^0(x^{\mu})}\mathrm{d}x^{\mu}$. The full expression of the asymptotic line element in Lyra geometry under the gauge function defined in eq.(\ref{T.2}) is:
\begin{equation}\label{Metric.3}
\mathrm{d}s^2=C^2exp(-\frac{4\sqrt{A}}{\alpha r^{\alpha/2}})[-exp(\frac{\frac{3}{4}A{\alpha}}{r^{\alpha}})\mathrm{d}t^2+exp(-{\frac{3}{4}A(\alpha^2-2)\frac{\alpha}{r^{\alpha}}})\mathrm{d}r^2+r^2\mathrm{d}\theta^2+r^2sin{\theta}^2 \mathrm{d}\phi^2]
\end{equation}

Firstly, we will be focusing on the the character of the first term in the expression of spacetime interval in eq.(\ref{Metric.3}). In general relativity, the frequency of photon satisfies: $\nu_E\sqrt{-g_{00}(r_E)}=\nu_R\sqrt{-g_{00}(r_R)}$, where $E$ and $R$ represents the emitter and the receiver of the photon and $r_E$ and $r_R$ are the radial coordinate of the emitter and the receiver. In Lyra geometry, we shall use the 'usual metric', then the relationship above becomes:  $\nu_Ex^0(r_E)\sqrt{-g_{00}(r_E)}=\nu_Rx^0(r_R)\sqrt{-g_{00}(r_R)}$. In the line element in eq.(\ref{Metric.3}), assuming that we emit a photon from flat spacetime zone to the singularity at $r=0$, the observer at $r=0^+$ will observe a frequency $\nu_R \to 0$. The energy of the photon will be depleted as it approaching to the singularity. Physically, this is corresponding to the result shows in eq.(\ref{B.7}).
Here, we could call the surface with $x^0(r)\sqrt{-g_{00}(r)}= \infty$ as 'backward infinite red-shift surface' in Lyra geometry.
We will discuss more about the physical property of the singularity at $r=0$ in next section.

\begin{figure}
\centering
\includegraphics[width=5in]{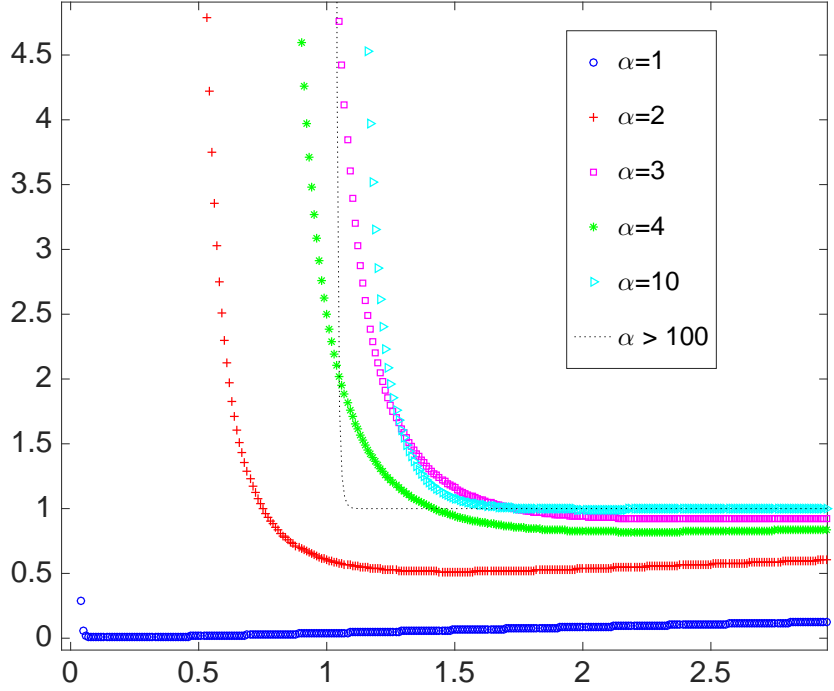}
\caption{(Color online.) Absolute value of the coefficient of $\mathrm{d}t^2$ in eq.(\ref{Metric.3}), $C^2exp(-\frac{4\sqrt{A}}{\alpha r^{\alpha/2}})exp(\frac{\frac{3}{4}A{\alpha}}{r^{\alpha}})$, $vs.$ radius coordinate $r$, where we choose $C=1$ and $A=1$. The plots are corresponding to $\alpha =1, 2, 3, 4, 10$. When $\alpha > 100$, the plot is asymptotically overlapping to the dashed line. We could observe that, when $\alpha$ is very large, the spacetime metric effectively have an surface with significantly large $x^0(r)\sqrt{-g_{00}(r)}$ near $r=1$. }
\label{Fig01}
\end{figure}
Moreover, we plotted the absolute value of the coefficient of the first term in the expression of spacetime interval in eq.(\ref{Metric.3}), $-(x^0)^2g_{00}=C^2exp(-\frac{4\sqrt{A}}{\alpha r^{\alpha/2}})exp(\frac{\frac{3}{4}A{\alpha}}{r^{\alpha}})$, versus the radial coordinate $r$ in Fig.(\ref{Fig01}).
We could observe that, $\alpha = 2$ corresponding to the plot of red cross in Fig.(\ref{Fig01}), the absolute value of the coefficient of $\mathrm{d}t^2$ increases rapidly when $r$ is around 0.5. 
When $\alpha$ is decreasing, the radius where this rapid climbing of $-(x^0)^2g_{00}$ happens will be more adjacent to $r=0$.
We could conclude that the surface with $x^0(r)\sqrt{-g_{00}(r)}= \infty$ is collapsing onto the singular point at $r=0$ from the mathematical expression even though we observe on the figure (\ref{Fig01}) that, for large $\alpha$, the 'backward infinite red-shift surface' is technically near $r=1$.
In the mean time, our gauge function $x^0(r)$ is still smooth at $r \in (0,\infty)$. 

Secondly, in the second term of the line element expression in eq.(\ref{Metric.3}), we have:
\begin{equation}\label{Metric.4}
(x^0)^2g_{11}=C^2exp(-{\frac{3}{4}A(\alpha^2-2)\frac{\alpha}{r^{\alpha}}}-\frac{4\sqrt{A}}{\alpha r^{\alpha/2}}).
\end{equation}
We will impose that $C=1$ here because as $r \to \infty$, $x^0(r)$ will be approaching to 1. This is corresponding to the situation that when we are far away from the singularity, the physics is the same as general relativity.
The definition of event horizon for a black hole is a null hypersurface in spacetime. For a static and spherical spacetime, the null hypersurface should also be static and spherical. Moreover, the normal vector $n^{\mu}$ of a hypersurface $f$ is a null vector. We will have:
\begin{equation}\label{Metric.5}
n_{\mu}n^{\mu}=0=g^{\mu\nu}n_{\mu}n_{\nu}=g^{\mu\nu}\frac{\partial{f}}{\partial{x^{\mu}}}\frac{\partial{f}}{\partial{x^{\nu}}}.
\end{equation}
In our spacetime interval (usual metric), eq.(\ref{Metric.5}) could be expressed as:
\begin{equation}\label{Metric.6}
(x^0)^{-2}g^{11}(\frac{\partial f}{\partial r})^2=0
\end{equation}
If the hypersurface $f$ is an event horizon existing in our model, $f$ should be irrelevant to the coordinates like: $t,\theta,\phi$. Here $ \frac{\partial f}{\partial r}$ should not be zero, otherwise $n_{\mu}$ will be zero. Eq.(\ref{Metric.6}) means that, if there is an event horizon existing in our model, the event horizon will be lying at: $(x^0)^{-2}g^{11} \to 0$. This is equivalent to: $(x^0)^{2}g_{11} \to \infty$. We could observe from eq.(\ref{Metric.4}) that, since $\alpha > 0$, there are three different situations:

1. $\alpha \in (0,\sqrt{2})$: $(x^0)^{2}g_{11}$ will be approaching to $+\infty$ as $r \to 0$.
The trend of $(x^0)^{2}g_{11}$ will be similar to the plot of the blue circle in Fig.(\ref{Fig01}). Under this condition, the event horizon $f$ will be collapsing onto $r=0$.

2. $\alpha=\sqrt{2}$: The behavior of $(x^0)^{2}g_{11}$ will be equivalent to $(x^0)^{2}$, since $g_{11}=1$ as $\alpha=\sqrt{2}$. Instead of divergence at $r = 0$, $(x^0)^{2}g_{11}$ will be approaching to zero when $r \to 0$. Therefore, no event horizon will be emerging in this scenario.

3. $\alpha>\sqrt{2}$: This situation is similar to situation 2. There will be no event horizon, since $(x^0)^{2}g_{11}$ is approaching to zero when $r \to 0$ and then the normal vector $n^{\mu}$ of the hypersurface has to be zero. Moreover, $(x^0)^{2}g_{11}$ will be approaching zero faster than the situation with $\alpha=\sqrt{2}$.

The analysis above has shown that in the spherical solution of Lyra's gravity with a smooth gauge function $x^0(r)=exp(-\frac{2\sqrt{A}}{\alpha r^{\alpha/2}})$ being imposed, when $\alpha \in [\sqrt{2}, \infty)$, the metric naturally created a 'naked singularity' as $r \to 0$. (This is unlike the disappearing event horizon in the Kerr metric of a rotating black hole, in which the event horizon disappear since the $r$ coordinate became complex under certain conditions.)
When $\alpha \in (0,\sqrt{2})$, there is an effective event horizon with very large $(x^0)^{2}g_{11}$ when $r \to 0^{+}$ even though $(x^0)^{2}g_{11}$ is technically divergent at $r=0$. 
In next section, we will further analysis the physical property of this spacetime solution. 
We will prove a physical scenario that a time-like curve which could reach arbitrarily small $r$ near the singularity at $r=0$ with finite integrated acceleration does NOT exist at all.

\subsection{A specific example at $\alpha=1$ with arbitrary $A$}

We could solve eq.(\ref{I.1}) when we choose the parameter $\alpha$ as $\alpha=1$. The asymptotic behavior of this solution is already plotted in Fig.(\ref{Fig01}) with the blue circle. The differential equation now becomes:
\begin{equation}\label{Sp.1}
        y^{{'}{'}}+\frac{2}{r}y^{'}+\frac{\frac{3}{4} A}{r^{2}}y^{'}=0
\end{equation}
General solution of this ODE is:
\begin{equation}\label{Sp.2}
        y=C_1+C_2 \cdot exp(\frac{3A}{4r}).
\end{equation}
where $C_1$ and $C_2$ are constants of integration. The correctness of this solution could be testified by plugging eq.(\ref{Sp.2}) back into eq.(\ref{Sp.1}).
Therefore, the time component of the line element would be like:
\begin{equation}\label{Sp.3}
\mathrm{d}s^2=C^2exp(-\frac{4\sqrt{A}}{r^{1/2}})[(C_1+C_2exp(\frac{\frac{3}{4}A}{r}))\mathrm{d}t^2 + ...
\end{equation}
where $C$ is the scaling constant from the gauge function $x^0(r)$. To set the integral constant $C_1$ and $C_2$, since physics will be the same as general relativity when $r \to \infty$, we have $C=1$, $C_1=0$ and $C_2=1$. 
Now we could write out the explicit expression of the line element as:
\begin{equation}\label{Sp.4}
\mathrm{d}s^2=exp(-\frac{4\sqrt{A}}{r^{1/2}})[(-exp(\frac{\frac{3}{4}A}{r}))\mathrm{d}t^2+exp({\frac{3}{4}\frac{A}{r}})\mathrm{d}r^2+r^2\mathrm{d}\theta^2+r^2sin{\theta}^2 \mathrm{d}\phi^2]
\end{equation}
This result agrees with the conclusion we have drawn above using the dominant balance method: When $r \to 0^{+}$, the metric component $-(x^0)^2g_{00}=exp(\frac{3A}{4r}-\frac{4\sqrt{A}}{r^{1/2}}) \to \infty$, which is in accordance with the asymptotic result.
Also, $(x^0)^{2}g_{11} =exp(\frac{3A}{4r}-\frac{4\sqrt{A}}{r^{1/2}}) \to \infty$, belongs to the situation 1, $\alpha \in (0,\sqrt{2})$, as we have mentioned above.

\section{Behavior Of Timelike Observer Under This Metric}
For a Schwarzschild black hole in general relativity under Riemannian geometry, when $r \to 0$, the time component of the metric goes to positive infinity:
\begin{equation}\label{B.1}
g_{00}=-(1-\frac{2M}{r}) \to +\infty
\end{equation}
Nevertheless, for our solution under the smooth gauge function $x^0$ defined in eq.(\ref{T.2}), we have the time component of the usual metric from eq.(\ref{Metric.3}), $(x^0)^2g_{00}$. When $r \to 0$:
\begin{equation}\label{B.2}
(x^0)^2g_{00}= -exp(\frac{\frac{3}{4}A{\alpha}}{r^{\alpha}}-\frac{4\sqrt{A}}{\alpha r^{\alpha/2}}) \to -\infty,
\end{equation}
where $A > 0$ and $\alpha > 0$. Here we choose $C$ in eq.(\ref{Metric.3}) to be 1. 

Physically, when $(x^0)^2g_{00} \to -\infty$, any certain observers, like a spaceship (corresponds to an arbitrary timelike curve), could not arrive at the singularity in the origin $r=0$. We could prove that the integrated acceleration must satisfy\cite{chakrabarti1983timelike,geroch1982singular,zheng1997thermodynamics}:
\begin{equation}\label{B.3}
\int^{\tau_{0}}_{\tau_{1}} a \mathrm{d}\tau \geq \int^{\tau_{0}}_{\tau_{1}} \mathrm{d}ln\sqrt{-(x^0)^2g_{00}}
\end{equation} 
where $\tau$ is the proper time which the spaceship measured, $\tau_{0}$ is the starting time of the spaceship and $\tau_{1}$ is the time when the spaceship arrives the singularity at $r=0$. $a$ is the norm of the acceleration defined in eq.(\ref{B.4}) below. 

Here we could observe that our spacetime following eq.(\ref{Metric.3}) is stationary. So we could define a timelike or null Killing vector field $t^{\mu}$ in our spacetime. Our timelike curve $\gamma$ is defined as the trajectory of the physical observer (spaceship). $\xi^{\mu}$ is the unit tangent vector with respect to $\gamma$. Moreover, we could introduce the effective energy $E$ as $E=-\xi^{\mu}t_{\mu}$. Furthermore, four-acceleration could be written as $a^{\nu}=\xi^{\mu}\nabla_{\mu}\xi^{\nu}$ and the projection operator could be written as: $h_{\mu\nu}=g_{\mu\nu}+\xi_{\mu}\xi_{\nu}$.

Step 1: To prove: $|\xi^{\mu}\nabla_{\mu}E| \leq aE$.
Proof: The left hand side: $|\xi^{\mu}\nabla_{\mu}E|=|-a_{\mu}t^{\mu}|=|a_{\mu}t_{\nu}h^{\mu\nu}|$. Since $h^{\mu\nu}$ is positive definite, we have: 
\begin{equation}\label{B.4}
|a_{\mu}t_{\nu}h^{\mu\nu}| \leq (h_{\mu\nu}a^{\mu}a^{\nu})^{\frac{1}{2}}(h_{\rho\sigma}t^{\rho}t^{\sigma})^{\frac{1}{2}}=a(t^{\mu}t_{\mu}+E^2)^{\frac{1}{2}}
\end{equation}
Follow the definition in Lyra geometry, $t^{\mu}t_{\mu}=(x^0)^2 g_{00}< 0$, we have: $|\xi^{\mu}\nabla_{\mu}E| \leq aE$. We then move the $E$ to the left hand side of $|\xi^{\mu}\nabla_{\mu}E| \leq aE$:
\begin{equation}\label{B.5}
a \geq \frac{ \mathrm{d}(lnE)}{\mathrm{d}\tau}
\end{equation}
since the unit tangent vector is defined as: $\xi^{\mu}=\frac{\mathrm{d}x^{\mu}}{\mathrm{d}\tau}$, where $\tau$ is the proper time on $\gamma$.

Step 2: To prove $E \geq \sqrt{-(x^0)^2 g_{00}}$.
Proof: Since $E=-\xi^{\mu}t_{\mu} \geq (-t^{\mu}t_{\mu})^{\frac{1}{2}}=\sqrt{-(x^0)^2 g_{00}}$.

Now we have proved the character of integrated acceleration in eq.(\ref{B.3}). In addition, Zhao\cite{zheng1997thermodynamics} proved that not just the integrated acceleration, the acceleration $a$ itself will also diverge when the observer approaches the singular region:
\begin{equation}\label{B.6}
\int^{\tau_{0}}_{\tau_{1}} a \mathrm{d}\tau = +\infty, \quad \mathrm{lim}_{\tau \to \tau_{0}} a= +\infty
\end{equation}

On the other hand, if we assume that the spaceship has a static mass $m_{0}$ at time $\tau_{1}$, when time $\tau \to \tau_{0}$, the mass of the spaceship is $m$. It can be proved\cite{chakrabarti1983timelike} that:
\begin{equation}\label{B.7}
    \int^{\tau_{0}}_{\tau_{1}} a \mathrm{d}\tau \leq -[\mathrm{ln}(m)-\mathrm{ln}(m_{0})].
\end{equation}
When the spaceship arrives the singularity, the left hand side diverge, this would demand $m \to 0$. Then at this moment the mass of this spaceship will decrease to zero. The spaceship will have zero static mass. Here, we could deduce that there is no timelike curve which could reach arbitrarily small $r$ value with finite integrated acceleration $a$ existed in this solution. This character of our solution is similar to Reissner-Nordstr\"{o}m metric\cite{reissner1916eigengravitation,nordstrom1918energy} in which the singular region of it is also inaccessible.

\section{Discussion And Outlook}
The naked singularity is existing or not plays a pivot role in cosmic censorship.
In this paper, we derived the usual metric of the spacetime in Lyra geometry under smooth gauge function. 
When $\alpha \in (0,\sqrt{2})$ in eq.(\ref{Metric.3}), the event horizon is collapsing onto the singularity at $r=0$. 
Moreover, when $\alpha \in [\sqrt{2}, \infty)$, the event horizon does NOT exist at all.
In the end, we also proved that no time like curve could reach $r=0$ with finite integrated acceleration. This means that the zone of the naked singularity is inaccessible physically.
Furthermore, with the explicit form of the 'naked singularity' presented in Lyra geometry, we could use Virbhadra-Ellis lens equation\cite{virbhadra2002gravitational} for the calculation of image position and for the image magnification in gravitational lensing effect. The singularity fabricated in our model just depends on the local behavior of the gauge function. If our gauge function is spherically symmetric with zero at the center of it as in eq.(\ref{T.2}), there will be a 'naked singularity' created mathematically.

\section{Acknowledgement}

The author wish to thank Qinghai Wang, Mengjiao Shi and Teng Zhang for helpful discussions. This work is supported in part by the Ministry Of Education in Singapore.

\appendix
\renewcommand{\appendixname}{Appendix~\Alph{section}}

\section{Using The Standard Frobenius Method}

In this appendix, we try to consider another possibility: What if the gauge function $x^0$ is not just a determined function and related to the $g_{00}$ component of the metric? Here we try to apply one specific gauge function to solve the ODE eq.(\ref{E.4}) using the standard Frobenius Method.

Firstly, we observe the ODE in equation (\ref{E.4}):
\begin{equation}\label{A.1}
y^{{'}{'}}+y^{'}(\frac{2}{r}+\frac{3}{4}rf(r))=0,
\end{equation}
where $y=e^{\nu(r)}=-g_{00}$ and $f(r)$ is defined as: $f(r)=[x^{0'}(r)/x^0(r)]^2$.
We would like to assume that $x^0$ is dimensionless, then the dimension of $f(r)$ is $-2$ in natural unit. 
Here we choose $f(r)= \frac{Ay}{y^{'}r^3}$, which could make the coefficient $A$ dimensionless. We put $f(r)= \frac{y}{y^{'}r^3}$ into the ODE (\ref{A.1}) and then we have:
\begin{equation}\label{A.2}
    y^{{'}{'}}+\frac{2}{r}y^{'}+\frac{3}{4}\frac{A}{r^2}y=0,
\end{equation}
This ODE with the power series expansion with respect to $r=0$ could be solved using standard Frobenius Method. The first solution to this ODE has the form:
\begin{equation}\label{A.3}
    y=x^{\alpha}\sum_{n=0}^{\infty}a_nx^n,
\end{equation}
where $a_0 \neq 0$. Plug this series into the equation (\ref{A.2}), we will have a series of equation with respect to the coefficients of the series in eq.(\ref{A.3}). $A$ is a positive real number. Here $\alpha_1$ and $\alpha_2$ are the two roots of the equation coupled with the coefficient $a_0$: $\alpha^2+\alpha+\frac{3}{4}A=0$.
The two roots are: 
\begin{equation}\label{A.4}
\alpha=\frac{-1 \pm \sqrt{1-3A}}{2}
\end{equation}
Different $A$ renders different $\alpha_1$ and $\alpha_2$, and different $\alpha_1$ and $\alpha_2$ correspond to different cases in Frobenius method\cite{bender2013advanced}.

Specifically, we solved the case when $A=1$, the calculated result is:
\begin{equation}\label{A.5}
y=-g_{00} = C_1\cdot cos[\frac{\sqrt{2}}{2}ln(r)]/\sqrt{r} + C_2 \cdot sin[\frac{\sqrt{2}}{2}ln(r)]/\sqrt{r}
\end{equation}
where $C_1$ and $C_2$ are constants of integration.
The  correctness  of  this  solution  could  be testified by plugging eq.(\ref{A.5}) back into eq.(\ref{A.2}). 
Since $sin[\frac{1}{2}\sqrt{2}ln(r)]$ and $cos[\frac{1}{2}\sqrt{2}ln(r)]$ are limited within $[-1,1]$ at all times, the behavior of them near $r \to 0^{+}$ and $r \to \infty$ is dominated by $\sqrt{r}$. When $r \to \infty$, we have $\sqrt{r} \to \infty$ and $g_{00} \to 0$ for all possible $C_1$ and $C_2$. In the mean time, as $r \to 0^{+}$, the $g_{00}$ is oscillating stronger and stronger until becomes divergent at $r = 0^{+}$. 


Furthermore, we impose that $C_1=0$ to make the calculation more neat. We take the first order derivative with respect to $r$ of function $y=C_2 \cdot sin[\frac{\sqrt{2}}{2}ln(r)]/\sqrt{r}$. Here $y=-g_{00}$. Then we have:
\begin{equation}\label{A.6}
y^{'}=C_2 \cdot cos[\frac{\sqrt{2}}{2}ln(r)]\frac{\sqrt{2}}{2}r^{-\frac{3}{2}}  -  C_2 \cdot sin[\frac{\sqrt{2}}{2}ln(r)]\frac{1}{2}r^{-\frac{3}{2}}
\end{equation}
Following the definition of $f(r)$ in eq.(\ref{A.1}), $f(r)=[x^{0'}(r)/x^0(r)]^2=\frac{y}{y^{'}r^3}$, we will have:
\begin{equation}\label{A.7}
\frac{x^{0'}(r)}{x^0(r)}=\sqrt{f(r)},
\end{equation}
in which the explicit form of $f(r)$ is:
\begin{equation}\label{A.8}
f(r)=\frac{2}{r^2(\sqrt{2} \cdot cot[\frac{\sqrt{2}}{2}ln(r)]-1)}
\end{equation}
Plug eq.(\ref{A.8}) back into eq.(\ref{A.7}), we will have:
\begin{equation}\label{A.9}
ln[x^0(r)]=\int \frac{\sqrt{2}}{r \cdot \sqrt{\sqrt{2} \cdot cot[\frac{\sqrt{2}}{2}ln(r)]-1}}\mathrm{d}r + C_3,
\end{equation}
where $C_3$ is the constant of integration. Now we will have the expression of the gauge function in this specific model as:
\begin{equation}\label{A.10}
x^0(r)=A_1 \cdot exp[\int \frac{\sqrt{2}}{r \cdot \sqrt{\sqrt{2} \cdot cot[\frac{\sqrt{2}}{2}ln(r)]-1}}\mathrm{d}r],
\end{equation}
where any constants of integration could be absorbed into the constant $A_1$. Next step, we will invoke eq.(\ref{component4}) to solve the $\lambda(r)$. Since $y=e^{\nu(r)}=-g_{00}$ and $y=C_2 \cdot sin[\frac{\sqrt{2}}{2}ln(r)]/\sqrt{r}$, we could calculate the derivative of $\nu(r)$ with respect to $r$:
\begin{equation}\label{A.11}
\nu^{'}(r)=cot(\frac{\sqrt{2}}{2}ln(r))\frac{\sqrt{2}}{2}\frac{1}{r}-\frac{1}{2r}
\end{equation}
Follow the ODE in eq.(\ref{component4}) and the expression of $f(r)$ in eq.(\ref{A.8}), we could get the expression for $\lambda^{'}(r)$:
\begin{equation}\label{A.12}
\lambda^{'}(r)=-\frac{3}{r(\sqrt{2} \cdot cot[\frac{\sqrt{2}}{2}ln(r)]-1)}-cot(\frac{\sqrt{2}}{2}ln(r))\frac{\sqrt{2}}{2}\frac{1}{r}+\frac{1}{2r}
\end{equation}
In the end, we will get the formula for $g_{11}$ component of the metric in this model, which is:
\begin{equation}\label{A.13}
g_{11}=e^{\lambda(r)}=B \cdot exp[- \int \frac{3}{r(\sqrt{2} \cdot cot[\frac{\sqrt{2}}{2}ln(r)]-1)}\mathrm{d}r - \int cot(\frac{\sqrt{2}}{2}ln(r))\frac{\sqrt{2}}{2}\frac{1}{r}\mathrm{d}r + \int \frac{1}{2r}\mathrm{d}r],
\end{equation}
where all the constants of integration are absorbed into the factor $B$. After we perform the last two indefinite integrals, we have:
\begin{equation}\label{A.14}
g_{11}=B^{'} \cdot exp[- \int \frac{3}{r(\sqrt{2} \cdot cot[\frac{\sqrt{2}}{2}ln(r)]-1)}\mathrm{d}r  - ln(sin(\frac{\sqrt{2}}{2}ln(r))) + \frac{1}{2}ln(r)]
\end{equation}
Now we have derived the expression of the spacetime interval (usual metric) in this model using Frobenius method with $C_1=0$ in eq.(\ref{A.5}). The usual metric is:
\begin{equation}\label{A.15}
\begin{split}
\mathrm{d}s^2=A^{'} \cdot exp[\int \frac{2\sqrt{2}}{r \cdot \sqrt{\sqrt{2} \cdot cot[\frac{\sqrt{2}}{2}ln(r)]-1}}\mathrm{d}r]\{ 
-C_2 \cdot sin[\frac{\sqrt{2}}{2}ln(r)]/\sqrt{r} \cdot \mathrm{d}t^2 \\+ B^{'} \cdot exp[- \int \frac{3}{r(\sqrt{2} \cdot cot[\frac{\sqrt{2}}{2}ln(r)]-1)}\mathrm{d}r  - ln(sin(\frac{\sqrt{2}}{2}ln(r))) + \frac{1}{2}ln(r)] \cdot \mathrm{d}r^2 \\ +r^2\mathrm{d}\theta^2+r^2sin{\theta}^2 \mathrm{d}\phi^2 \},
\end{split}
\end{equation}
where $A^{'}=A_1^2$. Here we have solved the usual metric in this model using Frobenius method under gauge function which is related to the $g_{00}$ component of the metric.

\end{document}